\documentclass[aps,prb,twocolumn,floatfix,superscriptaddress,nofootinbib]{revtex4-2}

\usepackage{amsmath,amssymb,bm}
\usepackage{graphicx}
\usepackage[hidelinks]{hyperref}
\usepackage{orcidlink}

\begin{document}
\setlength{\parindent}{0pt} 
\setlength{\parskip}{0.2em} 

\title{Quantum eigenvalues and eigenfunctions of an electron confined between conducting planes}

\author{Don MacMillen \orcidlink{0009-0001-2911-5210}}
\affiliation{San Mateo, California}
\email{don.macmillen@gmail.com}

\date{\today}

\begin{abstract}
  Two of the most iconic systems of quantum physics are the particle
  in a box and the Coulomb potential (the third is, of course, the
  harmonic oscillator). In this expository paper, we consider the
  quantum solution to the problem of an electron confined between the
  grounded planes of an infinite capacitor. The potential arises from
  the image charges that form in the grounded planes, along with the
  added condition that at x = 0, L, where L is the distance between
  the planes, the wavefunction must be zero. This effectively couples
  a hydrogen like system to a particle-in-a-box (PIB) based on L, the
  distance between the planes. The problem of finding the
  electrostatic potential of this infinite series of image charges is
  an old one, going back to at least 1929 \cite{Kellogg}. Here, we
  give a short derivation for one of the limiting cases that yields a
  compact expression that recovers the potential first presented in
  \cite{Barton} and show how the Kellogg infinite summation
  formula converges to that value. We note here that this potential is
  a symmetric double well potential, so there will be many familiar
  properties of its solutions. Then using that potential, we solve
  Schr\"odinger's equation using a spectral technique. The limiting
  forms of a particle in a box for small L (and high E), and that of a
  (degenerate) bound image charge \cite{Straub} \cite{Echen} at large
  L and small energy are recovered.  We also discuss the tunneling
  level splitting that occurs in the transition from the large L to
  the small L regime.
\end{abstract}


\maketitle

\section{Introduction}

We consider the quantum solution to the problem
of an electron confined between the grounded planes of an infinite
capacitor. To solve that quantum problem we need the potential of the
electron in that configuration. This electrostatic problem of finding
that potential is an old one. While it is not explicitly in Maxwell's
"A Treatise on Electricity and Magnetism" it is perhaps implicitly
contained in the problem of image charges in the presence of a
conducting sphere, when the radius is allowed to go to infinity \cite{Dick1}
\cite{Dick2}.

In 1929 Kellogg \cite{Kellogg} considers the problem of calculating the potential
at any location (x, y, z) in the space between two parallel conducting planes
where a unit charge is located at $(c,0,0)$ with $0<c<L$. Table~\ref{tab:convergence}
illustrates the convergence of the truncated image-charge series toward the closed-form
result in Eq.~(\ref{digamma}).

In a series of papers by different authors, and largely appearing in the
American Journal of Physics, the problem of computing the charge distribution
on the two planes is considered. Wong and Kittel \cite{Kittel} develop a series
solution to this problem, utilzing results from J.D. Jackson \cite{Jackson}. They
also develop a different series solution for the potential far from the charge
that has better convergence properties than the original Kellogg summation.

Pumplin \cite{Pumplin} obtains Green's function, again using the method of images,
and then uses a Sommerfeld-Watson transform to obtain an integral representation
for the potential. He then derives an asymptotic expression for the potential,
which is found to fall off exponentially. A very interesting result that 
shows that the two planes are very effective in screening a charge that is
placed between them.

Glasser \cite{Glasser} then simplifies Pumplin's results and obtains a closed form
solution along a line that passes directly through the charge. This is not
exactly the potential that we seek, but it is noteable in that it results
in an expression with two digamma functions.

Barton \cite{Barton} and Babiker and Barton \cite{Babiker} perform a quantum
electodymics calculation to get the radiative corrections to the energy of a
free and bound electron (hydrogen atom) in between the conducting planes. They
also produce the electrostatic energy of a free electron in between the planes
and Eqn. (2.21) of \cite{Barton} is identical to our Equ. (6) where we have
derived it using image charges and elementary methods. We don't consider the
radiative corrections futher here except to note that the leading term is
proportional to $\frac{1}{L^2}$ which reflects the boundary conditions on the
confining planes.

Interest in the problem of the potential and surface charge density in this
constrained system continued through to the present day \cite{Zahn} \cite{Pleines}
\cite{Simon} \cite{Samedov} and very recently \cite{Schmidt} \cite{Sousa}. 
Concepts developed by the early set of papers in the American Journal
of Physics continue to be relevant to problems in bilayer graphene
\cite{Gorbar} and in scanning quantum dot microscopy \cite{Wagner}. Further
applications of these ideas have played a role in image induced
surfaces states \cite{Cole} and in 2D materials \cite{Ando}. We expect that these
confined quantum systems to continue to be of interest and find new
applications.
    
The model that we extend here was first used for an electron bound by
its image potential to a single plane. There are two key
approximations used in the construction of this model.  The first is
that the electrons are excluded from the interior of the metal due to
Pauli repulsion from the electrons in the conduction band. This
approximation becomes less good as the energy increases beyond the
conduction band edge. The second approximation is that we assume that
the metal can instantainesly react to electron motion and perfectly
mirror the electon's charge. This is the infinite dielectric constant
approximation and is discussed in Jackson \cite{Jackson}.

This model of a single plane with a image potential induced bound state
is discussed further in Straub and Himpsel \cite{Straub} where they also
give spectrographic confirmation of the bound energy states.
This model has also been used for the physisorbtion of atomic hydrogen
onto the surface of a metal \cite{Bruch} \cite{MacMillen}.
More experimental realizations, along with many other aspects of these low
dimensional systems can be found in the book edited by Andrei \cite{Andrei}.

In order to move from the single conducting plane to the system of two
planes, we need the potential in this configuration (ie the early work
that we reviewed above).

In what follows, we first develop a closed form of the potential that
an electron experiences when confined between two conduction planes and
examine its form.  Then we utilize a spectral method to solve this
confined system to yield both the energies and states of the system.
Next, we present the results of the solves and discuss E as a
function of L, the approach of E(n) to $\frac{1}{2}(\frac{\pi n}{L})^2$
for small L and to $-\frac{1}{32 n^2}$ in the large L regime.

\section{The Image Potential}

We start with the series solution given by Kellogg on page 230 \cite{Kellogg}:

\begin{equation}
  \label{kellogg}
  \begin{split}
U = \sum_{n=-\infty}^\infty e ( \frac{1}{\sqrt{(x-2 n L - c)^2 + y^2 + z^2}} \\\\
    - \frac{1}{\sqrt{(x-2 n L + c)^2 + y^2 + z^2}}).
\end{split}
\end{equation}

This equation pairs image charges of opposite sign that differ by one
in their "generation".  The n=0 summand pairs the real charge with the
first generation (or primary) image charge to the left. The n=1
summand pairs the right primary image with the oppositely charged
image charge in the next generation (2nd) to the right. The positive index
sums all the image pairs to the right while the negative index sums
all the image pairs to the left.  As long as we keep this pairing of
the image charges intact, this series is, as noted by Kellogg,
absolutely and uniformly convergent.

Now we want the potential experienced by a real charge due only to the
image charges that it has induced and there are two important
modifications of this sum that we need to make. The first is that we
must drop part of the first term of the n=0 summand. (This would
correspond to the self-energy of the electron). The second is that we
must introduce a factor of $\frac{1}{2}$ because we are dealing with the
interaction of a charge with its own image (energy of assembly). This
is why the potential for a point charge in front of a single plane
conductor is $-\frac{q}{4x}$ and not $-\frac{q}{2x}$ \cite{Echen}.

We first consider the sum over the negative indices with $c = x$ and $y = z = 0$
and the additional factor of $\frac{1}{2}$ as just noted

\begin{equation}
\label{abskel}
\frac{1}{2} \sum_{n=-\infty}^{-1} (\frac{1}{|-2 n L|}  - \frac{1}{|2x-2 n L|} )
\end{equation}

which is easily seen to be equivalent to

\begin{equation}
\label{kel}
\frac{1}{4L} \sum_{n=1}^\infty ( \frac{1}{n} - \frac{1}{n - a} )
\end{equation}

where a = x / L so that $0 < a < 1$ which implies $|a - n| = n - a$. By
an easy manipulation of Equation 5.7.6 of the NIST Handbook \cite{NIST} we see
that this infinite sum is given in closed form by:

\begin{equation}
\sum_{n=1}^\infty (\frac{1}{n} - \frac{1}{n-a}) = \psi(1 - a) + \gamma
\end{equation}

where $\psi(z)(=\frac{\Gamma'(z)}{\Gamma(z)})$ is the digamma (or sometimes
the Psi) function and $\gamma$ is Euler-Mascheroni constant
(approximately 0.57721). By an analogous reasoning for the positive
index case of the Kellogg summation we find

\begin{equation}
\sum_{n=1}^\infty (\frac{1}{n} - \frac{1}{n+a}) = \psi(1 + a) + \gamma
\end{equation}

when we add back in the remaining part of the n=0 term (the primary image to
the left) and noting the
identity $\psi(x+1) - \psi(x) = \frac{1}{x}$ we reach our final
destination:

\begin{equation}
\label{digamma}
V(x)=\frac{1}{4L}\left[\psi(a)+\psi(1-a)+2\gamma\right],\qquad a\equiv \frac{x}{L}.
\end{equation}

Since $\psi(x) \sim - \frac{1}{x}$ for small x, we see immediately that this
has the correct limiting form for our problem.

We can also check the image modified Kellogg summation
formula against this result.  For $x = 0.5$ and $L = 1.0$ we find in
Table~\ref{tab:convergence} that even with 2000 terms we have only achieved 7
correct digits as compared to Eq.~(\ref{digamma}). So even though this series is
convergent, it is very slow to do so.

\begin{table}[tb]
\caption{\label{tab:convergence}Convergence of the truncated image-charge series toward the
closed-form result of Eq.~(\ref{digamma}).}
\begin{ruledtabular}
\begin{tabular}{lr}
Terms & Potential\\
\hline
20   & -0.6925809270\\
200  & -0.6931409927\\
2000 & -0.6931471181\\
Eq.~(\ref{digamma}) & -0.6931471806\\
\end{tabular}
\end{ruledtabular}
\end{table}

It is instructive to plot the potential of Eq.~(\ref{digamma}) on an interval alongside
the potential obtained from only the first generation of image charges, i.e.\
\begin{equation}
\label{eq:first-image}
V(x) = \frac{1}{4L}(-\frac{1}{a} - \frac{1}{1 - a})
\end{equation}

but it is even more interesting to also plot Eq. \eqref{eq:first-image}
vs. Eq.~(\ref{digamma}) when the constant $2\gamma$ has been left off
Eq.~(\ref{digamma})

\begin{figure}
\centering
\includegraphics[width=\columnwidth]{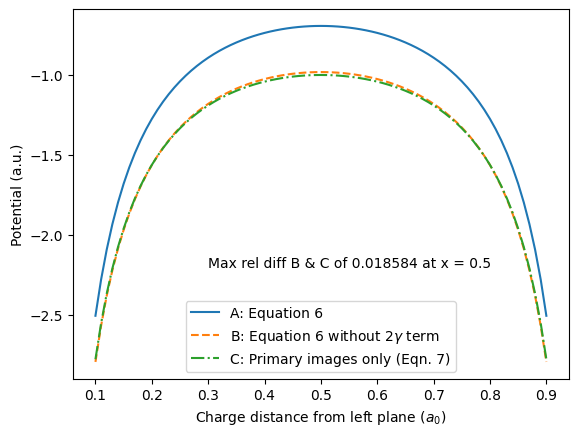}
\caption{\label{fig:potential}Comparison of Eq.~(6) with and without $2\gamma$ vs Eq.~(7). L = 1.0}
\end{figure}

We see from Fig.~\ref{fig:potential} that the effect on a real charge due to the
infinite series of its image charges in both directions, is largely to
add the constant $2\gamma$ to a potential made up of only the first
generation of images. (Note that the potential depends on ratio $a =
\frac{x}{L}$ so that form of the potential stays the same while the
maximum tends to zero as L gets large). This screening increases the
barrier between the two wells beyond that seen only by the first
generation of images. For low energies and large L, we would then
expect more charge localization near the walls when compared to only
the first generation image charges.

\section{Solving Schr\"odinger's equation with a spectral technique.}

In scaled units, the one-dimensional Hamiltonian is now seen to be
\begin{equation}
\label{eq:Hamilton}
H(x,L) = -\frac{1}{2}\frac{d^2}{d x^2} + \frac{1}{4L}\!\left(\psi(a)+\psi(1-a)+2\gamma\right)
\end{equation}

There are many excellent methods that can be used to solve this one
dimensional Schr\"odinger equation. Among the many we make note of two.
First, direct integration of the Schr\"odinger equation with the Numerov
method has a long history \cite{Cooley} and continues to be a popular choice 
\cite{Purev}. Second, starting in the 1980's, the Discrete Variable
Representation (DVR) has been, and continues to be, widely used for
many different problems \cite{Light} \cite{Shi}.

Here we will use a DVR-adjacent method known variously in the applied
mathematics literature as pseudospectral \cite{Boyd2003}, collocation,
discrete ordinates, selected points, or just simply as a spectral
method \cite{Boyd} \cite{Tref}.

In using a spectral method, the position variable is considered only
on a grid of suitably chosen points (which are usually connected to
various families of orthogonal polynomials \cite{Boyd} \cite{Tref}). Given
function values at these points, a spectral differentiation matrix can
be constructed, which becomes the core for the kinetic energy
operator. One of the key advantages for the spectral method is that
potential energy contribution is only needed on these grid points,
which are then added to the diagonal of the kinetic energy term to
form the Hamiltonian. This is in contrast to the Rayleigh-Ritz
procedure, where the matrix elements, which usually result in a dense
matrix, must be obtained by integration of the potential energy term
between two of the expansion's basis functions. One potential downside
to using the spectral method is that we no longer have any guarantees
that the calculated values are an upper bound on the true
value of the eigenvalue.

One of the challenges in applying a spectral method is in handling a
(semi-)infinite domain. One method is simply to truncate down to a
finite interval and examine the eigenvalues for convergence as the
interval is increased. This domain truncation can be problematic as
noted by Boyd \cite{Boyd2003}. Another approach is to use Laguerre or Hermite
polynomials that are defined over the semi-infinite(Laguerre) or
doubly infinite (Hermite) domains.  Another approach is to map or
scale the infinite interval to finite interval and then use standard
spectral methods. Here one must choose a scaling that works well for
both the problem and the chosen orthogonal polynomials (which are the
implied basis).

However, for this problem, we precisely want a truncated
domain. Moreover, since we have Coulomb-like behavior near each
boundary, a method that clusters the points near the endpoints of the
domain is likely to have better performance than one that does
not. This all suggests that the Chebyshev points (which have points
clustered near the endpoints proportional to $\frac{1}{\sqrt{1 - X^2}}$
as $\left|x\right| -> 1$ \cite{TrefApprox} and with the implied 'basis' of Chebyshev
polynomials) will have good performance. The final two pieces of the
puzzle are a simple linear scaling of the domain and ensuring that the
boundary conditions are satisfied. As shown in Chaper 7 of \cite{Tref},
homogeneous Dirichlet boundary conditions are easily enforced by using
only the interior Chebyshev points as well as truncating the second
order differentiation matrix. The Chebyshev grid points are given by
\begin{equation}
\label{eq:chebypts}
 x_j = \cos\!\left(\frac{j\pi}{M}\right),\quad j=0,1,\ldots,M.
\end{equation}

and, just as an example, the off diagonal elements of the differentiation
matrix are given by \cite{Tref}
\begin{equation}
\label{eq:chebydiff}
 (D_M)_{ij} = \frac{c_i}{c_j}\,\frac{(-1)^{i+j}}{x_i-x_j},\qquad i\neq j.
\end{equation}

where $i\neq j$, $i,j=0,1,\ldots,M$, and $c_i=2$ for $i=0,M$ (and $c_i=1$ otherwise).

Then the spectral Hamiltonian matrix that ensures the boundary conditions
are obeyed becomes

\begin{equation}
\label{eq:hamilmatrix}
 H_M(L)= -\frac{1}{2}\,D^{(2)}_{\mathrm{int}} + \operatorname{diag}\!\bigl(V(x_{\mathrm{int}},L)\bigr),
\end{equation}
where V refers to the potential of Eq.~(\ref{digamma}), the following dot
indicates that this potential is broadcasted over the truncated vector
of Chebyshev points, and the \texttt{diagm} constructs a diagonal matrix from
this vector (the actual code is different here for efficiency reasons,
see Appendix A). Here we see the great advantage of a spectral method
over a Rayleigh-Ritz approach, as noted earlier, we only need to
evaluate the potential on a vector of Chebyshev points. Finally, we
use the Julia \cite{Julia} \texttt{eigen} function to calculate the eigenvalues and
eigenvectors. By default, Julia uses the LAPACK solvers and specific
algorithms can be specified, if desired.

Taking all of these considerations into account results in a remarkably
compact program of under 40 lines of Julia code, the entirety of
which is reproduced in Appendix A.  Since we know what the limiting cases
should be, we can easily check that at very large separation we obtain
the single plane energy levels for a bound image charge of $-\frac{1}{32 n^2}$
\cite{Straub} only now with a (near) degeneracy of 2. Running the program
of Appendix A on an Intel Core i9-9980HK took 2.022 seconds and
produced the results of Table~\ref{tab:large_sep}.

\begin{table}[t]
\caption{\label{tab:large_sep}Energy levels at large separation ($L=10000$).}
\begin{ruledtabular}
\setlength{\tabcolsep}{2pt} 
\begin{tabular}{rr}
Calculated & Limiting \\
\hline
-0.0312500000288\textit{4847}   & -0.03125 \\
-0.0312500000288\textit{3924}  & -0.03125 \\
-0.00781250040389\textit{4916}  & -0.0078125 \\
-0.00781250040389\textit{3448}  & -0.0078125 \\
-0.003472224212910\textit{9305} & -0.0034722 \\
-0.003472224212910\textit{8464} & -0.0034722 \\
-0.001953131232235\textit{936}  & -0.001953125 \\
-0.001953131232235\textit{714}  & -0.001953125 \\
-0.001250015150416\textit{800} & -0.00125 \\
-0.001250015150416\textit{723}  & -0.00125 \\
\end{tabular}
\end{ruledtabular}
\end{table}

\section{Results and discussion}

In the small L regime, where the PIB-like solutions
dominate, the eigenvalues are still close to the ideal PIB values of
\begin{equation}
\label{eq:pib}
E_N(L) = \frac{1}{2}\left(\frac{\pi N}{L}\right)^2
\end{equation}

as can be seen in Table~\ref{tab:low_eigen}

\begin{table}[tb]
\caption{\label{tab:low_eigen}Energy levels at $L=1.0$ ($M=100$) and the corresponding quantum defect.}
\begin{ruledtabular}
\begin{tabular}{rrr}
$N$ & Calculated & Quantum defect\\
\hline
1 & 4.0122415062 & 0.0983071\\
2 & 18.467338037 & 0.0655065\\
3 & 42.940196396 & 0.050169\\
4 & 77.341129458 & 0.0411378\\
5 & 121.64356035 & 0.0351094\\
6 & 175.83576914 & 0.0307642\\
7 & 239.91151708 & 0.0274655\\
8 & 313.86707978 & 0.0248656\\
9 & 397.70005311 & 0.0227576\\
10 & 491.40879397 & 0.02101\\
\end{tabular}
\end{ruledtabular}
\end{table}

Here we show the quantum number (N), the calculated energy for L=1.0,
and the \emph{quantum defect}, that is, the value to subtract from N in
order to get the correct energy by using Eq. \eqref{eq:pib}. Note that the
quantum defects start out at $\sim 10\%$ but quickly decreases to $\sim 0.2\%$ at
N = 10.

\begin{figure}
\centering
\includegraphics[width=\columnwidth]{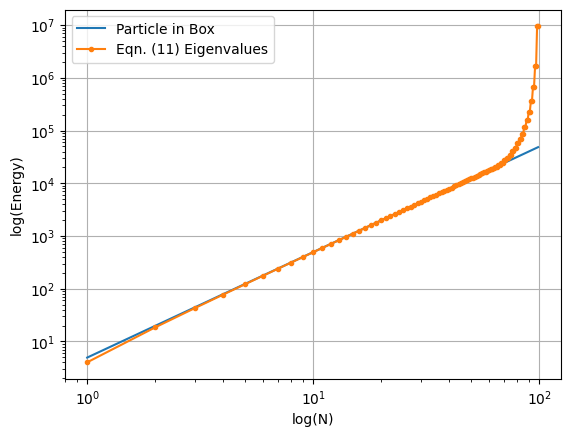}
\caption{\label{fig:large_n}Log(E) verses Log(N) for L=1.0, M=100}
\end{figure}

The visualization of the large N behavior of the energy is done best
using a log-log plot of N vs. E. Because of the quadratic dependence
on N we see a straight line graph with a slope of 2 in Fig.~\ref{fig:large_n}. At
small N, the calculated values are seen to differ from the PIB values.
At intermediate N, the two energy curves are very close together. At
large N we see the eigenvalues of the Hamiltonian matrix
\eqref{eq:hamilmatrix} rise steeply away from the PIB line and we know that
these eigenvalues are not indicative of the solutions to the partial
differential equation \eqref{eq:Hamilton}.

\begin{figure}
\centering
\includegraphics[width=\columnwidth]{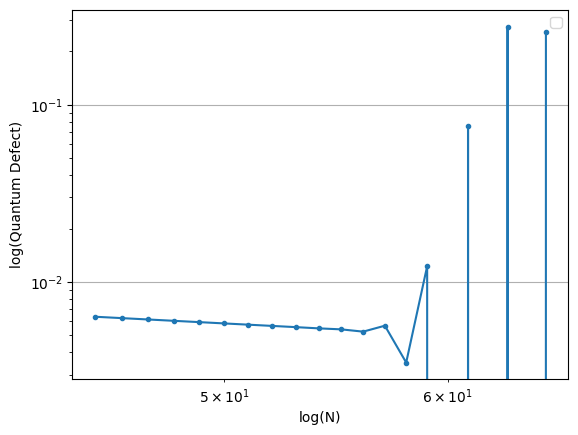}
\caption{\label{fig:log_defect}Log(quantum defect) vs log(N) for N=45:65, L=1.0, M=100}
\end{figure}

Looking at Fig.~\ref{fig:large_n} we might conclude that the eigenvalues are O.K.
up to about N=70, but that is not the case. A closer look at the
log(quantum defect) vs. log(N) in Fig.~\ref{fig:log_defect} reveals that after
gradually decreasing, it starts to fluctate wildly starting around N=55.

This is inline with the rule-of-thumb given in \cite{Boyd2003} where they
note that only $~\frac{M}{2}$ of the eigenvalues of the Hamiltonian
matrix \eqref{eq:hamilmatrix} will be a good approximation to the solutions
of \eqref{eq:Hamilton}. This general behavior remains true as the number of
Chebyshev grid points, M, is increased.

\begin{figure}
\centering
\includegraphics[width=\columnwidth]{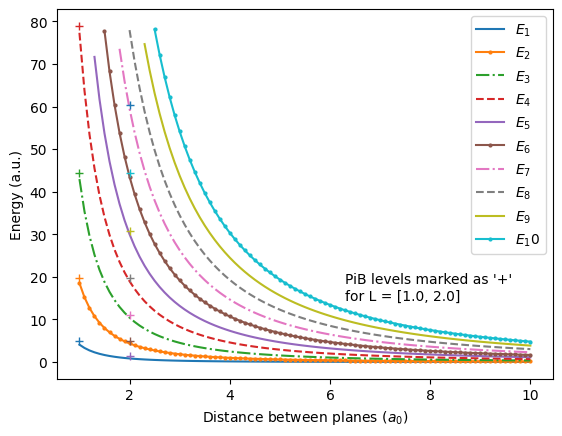}
\caption{\label{fig:ten_levels}Energy of the first 10 eigenvalues as function of distance L}
\end{figure}

Turning attention now to the L dependence in the small L regime, we
see the $\frac{1}{L^2}$ dependence in Fig.~\ref{fig:ten_levels}. There, the exact
PIB eigenvalues are marked for the first few levels at
L = 1,2 with a $+$. We see the steep fall off in energy as L grows
larger.

\begin{figure}
\centering
\includegraphics[width=\columnwidth]{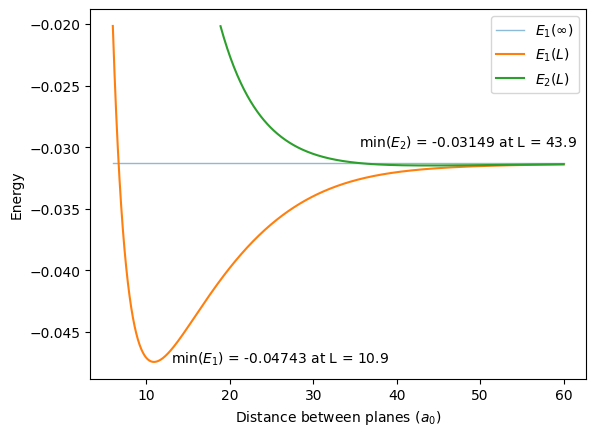}
\caption{\label{fig:first_pair}Energy splitting of first eigenvalue pair as function of distance L}
\end{figure}

If we zoom in on the first two eigenvalues, we see the level splitting
as L is varied, as shown in Fig.~\ref{fig:first_pair}. As L is
decreased from a large value, both of these levels decrease in energy
due to the image effects of both planes being closer.
Unsurprisingly, this is very similar to the level splitting observed
in the molecular ion $H_2^+$ \cite{Turbiner}.

The value of the energy splitting can be approximated analytically \cite{Herring} \cite{Garg}
where they note that this splitting is due to the tunneling through the central
barrier. We start with Eq (2.1) from \cite{Garg}
\begin{equation}
\label{eq:split}
\Delta E(L) = 2 \Psi_0 (\frac{L}{2}) \Psi_0 '(\frac{L}{2})
\end{equation}

Here, $\Psi_0$ refers to the ground state of an electron bound to a single plane.
This normalized wavefunction (in a.u.) is given by

\begin{equation}
\Psi_0(x) = \frac{x}{4} exp(-\frac{x}{4})
\end{equation}

In \eqref{eq:split} we have mapped the coordinates to the symmetry point of
the interval.  Since this splitting is due to tunnelling it makes
intuitive sense that it depends on the value of the wavefunction and
its derivative at the maximum height of the barrier.

Now \eqref{eq:split} is easily seen to be
\begin{equation}
\label{eq:firstsplit}
\Delta E(L) = \frac{L}{16} exp(-\frac{L}{4}) (1 - \frac{L}{8})
\end{equation}

\begin{figure}
\centering
\includegraphics[width=\columnwidth]{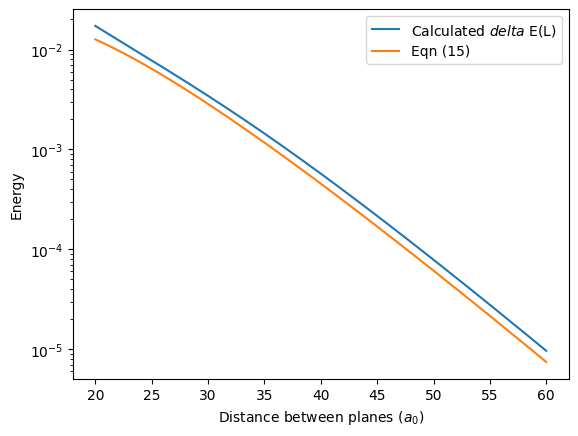}
\caption{\label{fig:garg_splitting}Delta E for first two states vs. Eq. \eqref{eq:firstsplit}}
\end{figure}

The prediction of \eqref{eq:firstsplit} verses the calculated results are
shown in a semilog plot in Fig.~\ref{fig:garg_splitting} where it appears that we are
only off by a small constant factor.

\begin{figure*}
\centering
\includegraphics[width=\textwidth]{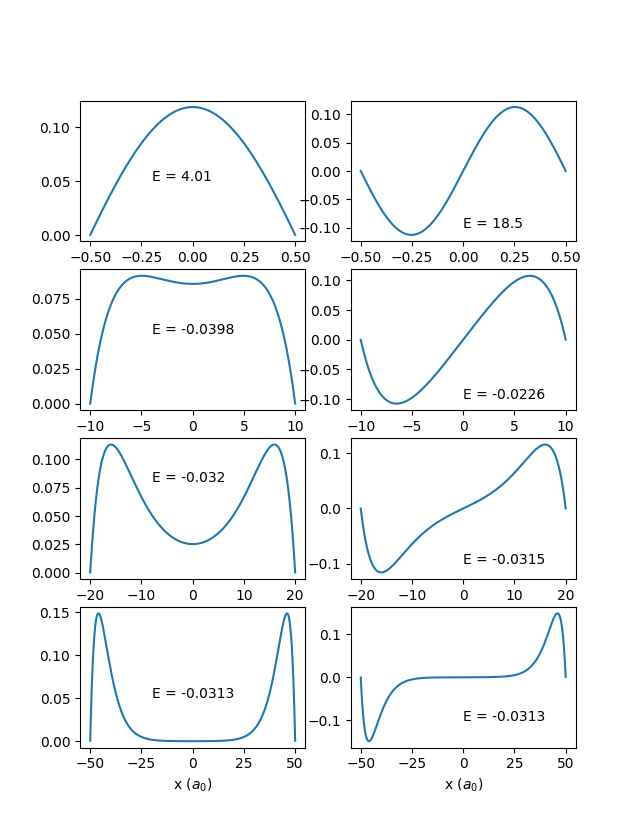}
\caption{\label{fig:waveforms}Waveform evolution as function of distance L for the first two eigenstates  \ Rows: L = [1.0, 20.0, 40.0, 100.0] \ First column: ground state, Second column 1st excited state. \ Note that the distance scales change as you go down the figure}
\end{figure*}

\begin{figure*}
\centering
\includegraphics[width=\textwidth]{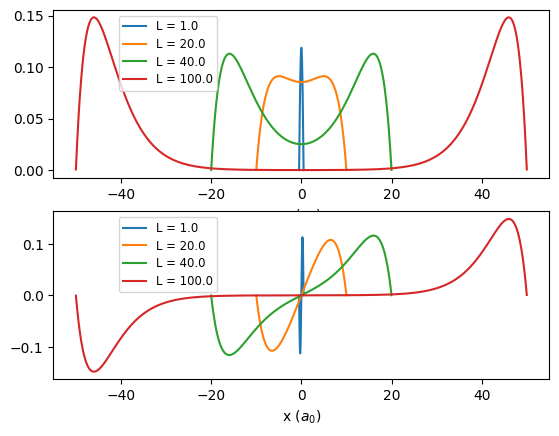}
\caption{\label{fig:mergedw}Same information as Fig.~\ref{fig:waveforms} but plotted on common x axis. Top graphis the ground state while the bottom graph is the first excited state}
\end{figure*}

As the distance between the two planes is increased, the ground state
wavefunction transforms from the essentially PIB ground state, where
the amplitude is concentrated in the center of the range, to one where
the amplitude is concentrated near the planes.  Fig.~\ref{fig:waveforms} shows the
evolution of the first two eigenfunction from the PIB solutions into
the symmetric and anti-symmetric combinations of the ground states of
two isolated electrons bound by their images. Fig.~\ref{fig:mergedw} shows
the same information as Fig.~\ref{fig:waveforms}, but now they are plotted on a
common x axis.

We note here that as the distance increases much after 100 $a_0$ the energy
splitting becomes very small and the returned eigenfunctions for the first
two states will be an arbitrary pair of orthogonal functions so that they
no longer possess the required even / odd symmetry. However the calculated
energy should still remain accurate.

\section{Conclusion}

We have only given a brief outline of the solution and properties of
this interesting model problem. We have not fully explained all of the
pertinant phenomina and issues and instead rely on the references for
much of the deeper theory, which is hoped to be a good guide for the
multiple topics touched on here.

Because of the advances in methods (spectral collocation) and software
tools (Julia, linear algebra solvers, Matplotlib, etc), this problem
becomes accessible to capable undergraduates. Further studies could
include the radiative corrections from \cite{Barton} and \cite{Babiker},
examining the splitting of higher excited states, and at large
constant L, understanding the transition from an image state to a PIB
state as the quantum number increases. Also, calculating and
understanding the induced charge density in the planes as a function
of L and the quantum number would be interesting. Additionally, using the
'quantum defect' as in Fig.~\ref{fig:log_defect} to assess the
numerical soundness of the calculated spectrum is appealing because
it does not depend on redoing the calculation with increasing number
of collocation points. Is there a way to generalize this proceedure?

Instead of trying to solve the current problem more correctly (ie
adding the radiative corrections as mentioned above) can we think of
different problem conditions or geometeries that lead to interesting
problems but still enjoy the simplicity and convergence of the current
methods?

One such idea would be to add an electric field between the planes
which would correspond to a charged capacitor. This would be
equivalent to adding a linear potential between the planes. In the
absence of the image charges, we know that the solution to the
Schrodinger equation is an Airy function. What happens when the image
charges are added? This would be a simple, almost one line change to
the current code.

We might also consider a different geometry where we are interested in
a point charge confined inside a co-axial cable between the core and
the sheathing where both of them are grounded. The axial symmerty would
suggest the use of cylindrical coordinates.

This would again result in an infinte series of image charges that would
need to be summed. Can we find a closed form expression for this sum?
Or has this problem already been solved? Are there any subtle quantum
symmetry issues when solving this system?

Finally, think of your own project, take the code, and run it for yourself!

\section*{Conflict of Interest Statement}
The authors declare that they have no conflict of interest related to this work.

\section{Appendix A}

As noted by Trefethen in \cite{Tref}, the calculation of D2 (the second order
differentiation matrix) by squaring D1 "is not the most stable method,
nor the most efficient." We can, for example, use the explicit formulas
for D2 as shown in \cite{Ehren} which reduces the $N^3$ complexity to $N^2$.
We can also avoid much of the domain scaling that is in the code below by
a more intelligent factoring. We have not done that in this code so that
it remains simple and easy to match to the discussion. What is remarkable,
is that it still gives excellent results.

\begin{verbatim}
using Base.MathConstants
using LinearAlgebra
using SpecialFunctions

# Modifed from Trefethen's
# "Spectral Methods in Matlab"
function cheb(M)
    M == 0 && return [[]], []
    # M+1 points. Use shifted sine for
    # symmetry. See Exercise 2.3 of
    # Trefethen's ATAP
    x = sinpi.((M:-2:-M) ./ 2M) 
    c = ones(M+1)
    c[[1,end]] .= 2.0
    c[2:2:end] .*= -1
    X = repeat(x, 1, M+1)
    dX = X - X'         
    D  = (c * (1 ./ c)') ./ (dX + I)
    D[diagind(D)] .-= sum(D, dims=2)
    return D, x
end

# digamma potential on the interval
#    [-L/2, L/2]
function doubleD(x, L)
    a = x / L + 0.5
    return (digamma(a) + digamma(1 - a) +
            2*eulergamma) / 4L
end

# Construct the Hamiltonian
function h_cheb_digamma(;M=10, L=10.0)
    D, x = cheb(M)
    # scale D and x to [-L/2, L/2]
    D .*= (2.0 / L)
    x .*= (L / 2)
    D2 = D^2
    # Set boundary conditions to
    # zero by truncation
    D2 = D2[2:M, 2:M]   
    H = -D2 / 2
    H[diagind(H)] .+= doubleD.(x[2:end-1], (L,))
    return H
end

# Check a limiting behavior...
function main()
    vals, vecs = eigen(
       h_cheb_digamma(M=2000, L=10000))
    img_exact = -(1 ./ (1:5) .^ 2) / 32
    for i in 1:10
        println(real(vals[i]), "  ",
                img_exact[div(i+1,2)])
    end
end
\end{verbatim}

\bibliography{refs}

\end{document}